\documentclass[]{mn2e}  

\newcommand{\diff}{\mbox{d}}
\usepackage{graphicx} 
\title[TRGB distances to composite stellar populations]{Tip of 
the Red Giant Branch distances to galaxies with composite stellar populations}
\author[M. Salaris \& L. Girardi]
       {Maurizio Salaris$^{1,2}$ \& L\'eo Girardi$^{3}$\\
$^1$Astrophysics Research Institute, Liverpool John Moores
        University, Twelve Quays House, Egerton Wharf, Birkenhead CH41 1LD,
        UK \\
$^2$Max-Planck-Institut f\"ur Astrophysik, Karl-Schwarzschild-Strasse
1, 85758 Garching, Germany\\
$^3$Osservatorio Astronomico di Trieste, INAF,
    Via Tiepolo 11, I-34131 Trieste, Italy}

\begin{document} 
 
 
\pagerange{\pageref{firstpage}--\pageref{lastpage}} \pubyear{2004} 
 
\maketitle 
 
\label{firstpage} 
 
\begin{abstract} 
The accurate determination of galaxy relative distances is extremely
important for the empirical calibration of  
the uncertain metallicity dependence of some standard candles like
Cepheid stars, or for studying the galaxy space distribution and
peculiar velocities. 
Here we have investigated the reliability of 
the widely used $I$-band Tip of the Red Giant Branch (TRGB) 
relative distances for a sample of Local Group galaxies with complex Star
Formation Histories (SFRs) and Age Metallicity Relationships (AMRs) 
namely the LMC, SMC and LGS3. The use of the $K$-band is also discussed.
By employing theoretical stellar population synthesis techniques, 
we find that using actual determinations of SFR and AMR of the LMC and
SMC, their RGB is populated by stars much younger (by $\sim$9~Gyr) 
than the Galactic globular cluster counterparts, on which the
$I$-band (and $K$-band) TRGB absolute magnitude is calibrated. 
This age difference induces a
bias in both the photometric metallicity estimates based on the comparison of
RGB colours with globular cluster ones, and the TRGB distances.
The extent of the distance bias -- which is not influenced by the 
actual value of the TRGB absolute magnitude zero point -- 
is strongly dependent on the specific 
TRGB technique applied, and on the assumed $I$-band bolometric
correction ($BC_I$) scale adopted; the correction to apply to the SMC--LMC 
distance modulus ranges from 0 mag up to +0.10 mag. 
LGS3 is an example of galaxy populated mainly by
old stars, so that photometric metallicity and distance estimates
using globular cluster calibrations are reliable. However, the relative
distance moduli between Magellanic Clouds and LGS3 are affected by 
the population
effects discussed for the LMC and SMC. The correction to apply to the
LGS3--LMC distance modulus ranges between $-0.05$ mag and to $+0.14$ mag, whereas in case
of the LGS3--SMC distance modulus it goes from $-$0.07 mag to $+$0.04 mag. 
In case of all three relative distances discussed before, the
correction to apply to the $K$-band TRGB distances are larger than the $I$-band case.

Our results clearly show that the presence of a well developed RGB
in the Colour Magnitude Diagram of a stellar system with a complex SFR does
not guarantee that it is populated by globular cluster-like red
giants, and therefore the TRGB method for distance determination 
has to be applied with caution. 
A definitive assessment of the appropriate corrections for
population effects on TRGB distances has however to wait for a
substantial reduction in the uncertainties on the $BC_I$
scale for cold stars.

\end{abstract}  
 
 
\section{Introduction} 

The red giant branch (RGB) is one of the most prominent and well
populated features in Colour-Magnitude-Diagram (CMD) of stellar
populations with ages larger than $\sim 1.5 - 2.0$ Gyr. The upper
boundary of the RGB locus is the so called tip of
the RGB (TRGB), populated by stars whose electron degenerate core
is to start the helium flash. At the helium flash the electron
degeneracy is lifted and quiescent helium burning will proceed in a non
degenerate core. The helium flash phase is extremely short,
and after the degeneracy is lifted, stars are located at much
lower luminosities during the quiescent central helium
burning. The subsequent asymptotic giant branch
phase -- that basically overlaps with the upper RGB and then reaches
higher luminosities -- is much faster than the RGB one, and therefore 
the TRGB manifests itself as a 
discontinuity in the differential luminosity function of bright red stars
in the CMDs of intermediate-age to old stellar populations. 

As for the TRGB absolute magnitude, 
theoretical and semiempirical works (see, e.g., Da Costa \&
Armandroff~1990, LFM93, Salaris \& Cassisi~1997, 1998) all
agree that observations of the TRGB in the $I$-band minimize its
dependence on age and chemical composition, so that the $M_I^{\rm TRGB}$ is
considered to be a very good standard candle for stellar populations
displaying a RGB in their CMDs. LFM93  presented a comprehensive 
calibration of the TRGB 
method to determine galaxy distances, using results 
from Da Costa \& Armandroff~(1990 -- hereafter DA90). 
The TRGB has been employed to derive distances to objects
in the Local Group and beyond, out to the Leo~I group and Virgo
(e.g. Lee, Freedman \& Madore~1993 -- hereinafter LFM93, Sakai et al.~1997, 
Harris et al.~1998, Cioni et al.~2000, Jerjen \& Rejkuba~2001, M\'endez
et al.~2002, Karachentsev et al.~2003, McConnachie et al.~2004). 
Such distance estimates are
very important for calibrating and testing other distance indicators, 
for measuring the
local velocity fields and also to determine star formation histories
of the galaxies under scrutiny. 

Due to the popularity of the TRGB as standard candle, 
a series of works have been devoted to improve various 
aspects of the TRGB method. Different techniques for the 
TRGB detection have been put forward,  
often to improve its detection in case of sparsely populated CMDs, 
or in presence of high photometric errors.
The most used method is the one presented by LFM93,
which employs an edge detection algorithm (zero sum Sobel kernel [-2, 0, +2]) 
to detect a discontinuity in the binned
luminosity function of bright RGB stars. A refinement of this method   
was presented in Sakai, Madore \& Freedman~(1996), who applied 
a continuous edge detection function to the Gaussian smoothed 
luminosity function.
Alternative techniques have been discussed by 
Cioni et al.~(2000), Sarajedini et al.~(2002), M\'endez et al.~(2002), 
Frayn \& Gilmore~(2003), McConnachie et al.~(2004). 

On the other hand, various theoretical, 
semiempirical or empirical calibrations of
$M_I^{\rm TRGB}$ have been published by Salaris \& Cassisi~(1997, 1998), DA90, 
Bellazzini et al.~(2001, 2004).
Calibrations of the absolute magnitude of the TRGB in infrared
bands ($J, H, K$) are also available (Ferraro et al.~2000, Bellazzini
et al.~2004), which largely avoid uncertainties 
related to the effect of extinction but, as we will see later, 
have a stronger dependence on the metallicity, and hence are likely more
sensitive to population effects.

Once a given TRGB calibration is employed  
-- the zero points of the existing calibrations span a range of $\sim$0.2
mag -- relative distance moduli obtained from the TRGB are considered to be
extremely precise, sometimes with error bars of the order of only 0.05
mag or less (e.g. McConnachie et al.~2004). 
This precision has prompted the use of
the TRGB as a reference standard candle on which to calibrate
empirically the metallicity dependence of other widely used stellar
distance indicators, like Red Clump stars (i.e. Pietrzy\'nski et
al.~2003) or Cepheid variables (i.e., Sakai et al.~2004).

A fundamental point to be stressed is that 
all calibrations of $M_I^{\rm TRGB}$ (or the infrared counterparts)
applied to determine TRGB galaxy distances assume that
the observed stellar populations are 'old', with ages of the order of
the Galactic globular cluster ones (12-14 Gyr). This stems from the
implicit assumption that a well developed RGB is populated by old,
globular cluster-like stars. This is possibly a very good assumption
when observing, e.g., haloes of spiral galaxies, but it is not
justified in case of composite stellar populations. Barker, Sarajedini
\& Harris~(2004) have recently studied the case of composite stellar
populations with a fraction of stars with ages of the order of 1.5
Gyr, corresponding to the RGB phase transition (Sweigart, Greggio
\& Renzini~1990), i.e., when the He
ignition happens in a progressively less degenerate
core. During this transition the TRGB moves at lower
luminosities with respect to older populations and Barker et
al.~(2004) found that if more than 30\% of the total number of stars 
created in a given stellar population have an age of this
order, the distance obtained applying the standard TRGB calibrations
is overestimated by $\sim 10 - 20$\%. However, the authors conclude
that the galaxy distances obtained from the TRGB method up to now have not
been affected by this effect, since none of the observed 
stellar populations appears to have a star formation history of this kind.

In this paper we complement the analysis by Barker et al.~(2004) by
studying the effect of the star formation history on the TRGB
distances, considering specific examples of galaxies in the Local Group,
e.g. LMC, SMC, and LGS3; our results are independent of both the TRGB 
zero point calibration, and the adopted technique to detect the TRGB from the 
stellar luminosity function. We will show how even without the
presence of a substantial population at the RGB phase transition,
systematic errors of 0.1--0.2 mag in TRGB relative distances might be
present when the stellar population under scrutiny is not uniformly as
old as the Galactic globular clusters.
The paper is organized as follows. Section~2 discusses briefly the TRGB
method and its calibration in both $I$ and infrared passbands, while
Sect.~3 presents a detailed reanalysis of the TRGB distances to LMC, SMC and
LGS3 and discusses the biases in their relative distances if one
follows the traditional use of the TRGB method. 
Conclusions follow in Sect.~4.

\section[]{TRGB brightness as a function of age and metallicity} 

\begin{figure}
\includegraphics[width=8.3cm]{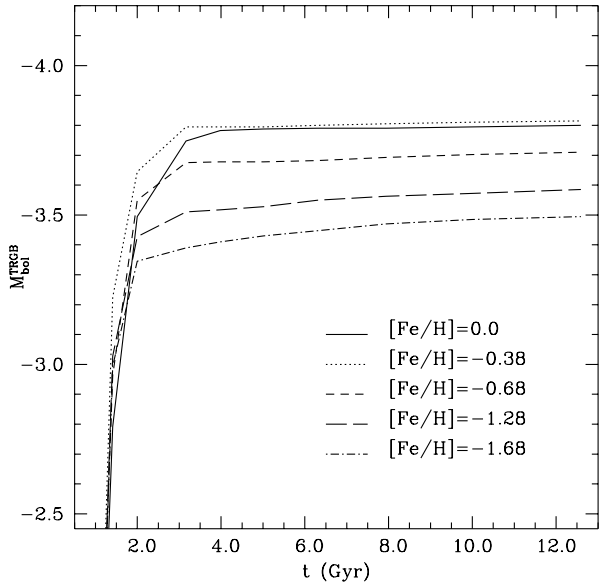}
\caption{$M_{\rm bol}$ of the TRGB as a function of age and [Fe/H],
from the theoretical models by Girardi et al.~(2000)}
\label{bolTRGB}
\end{figure}

The bolometric luminosity of TRGB stars as derived from stellar
evolution models is largely 
determined by the size of the He core at the He flash,
once the initial chemical composition is fixed\footnote{At a fixed initial
composition, $M_{\rm bol}^{\rm TRGB}$ also presents a very small dependence on the
envelope composition changes caused by the first dredge-up. This dependence
however is much smaller than the one on the core mass.}. Since low mass stars 
ignite He all with similar core masses (slightly increasing
for decreasing stellar mass), $M_{\rm bol}^{\rm TRGB}$ changes by a few
hundredths of magnitudes in the typical age range of Galactic globular
clusters (ages between $\sim$ 8 and $\sim$ 13 Gyr), as shown in
Fig.~\ref{bolTRGB} using the scaled-solar models by Girardi et al.~(2000).
When approaching the RGB phase transition $M_{\rm bol}^{\rm TRGB}$ increases sharply,
due to the lifting of electron degeneracy in the He core, that causes
He ignition to occur at significantly smaller core masses.

On the other hand, once the age is fixed, $M_{\rm bol}^{\rm TRGB}$ 
decreases for increasing metallicity, in spite of the decrease of the 
He core mass at the TRGB. As shown, e.g., in Kippenhahn \&
Weigert~(1990) on the basis of homology relations, this property stems from the
fact that at a given core mass an increase of the molecular weight in the
H-burning shell causes an increase of the RGB star luminosity.
The net effect is a decrease of the TRGB brightness with decreasing [Fe/H];
in general, $M_{\rm bol}^{\rm TRGB} \propto -$0.20 [Fe/H] at a fixed age far
from the RGB phase transition (say $\ga 3$ Gyr) and for metallicities 
well below the solar value.

DA90 have provided a much used semiempirical calibration of 
$M_{\rm bol}^{\rm TRGB}$ as a function of [Fe/H] in the Galactic globular
cluster regime (old ages and [Fe/H] at most $\sim -$0.7). 
They employed empirical determinations of the bolometric
magnitude of RGB stars in a sample of clusters of known metallicity,
assumed a globular cluster distance scale and fixed the 
slope of the $M_{\rm bol}^{\rm TRGB}$ vs. [Fe/H] relationship to the value
given by Sweigart \& Gross~(1978) theoretical models, obtaining
\begin{equation}
M_{\rm bol}^{\rm TRGB}=-0.19 \ [Fe/H] - 3.81
\label{eq1}
\end{equation}

A previous semiempirical determination by Frogel et al.~(1983), as
well as  more recent theoretical and semiempirical evaluations of 
this relationship
(e.g. Salaris \& Cassisi~1997, 1998, Ferraro et al.~2000) including 
the results displayed in Fig.~\ref{bolTRGB}, have a slope very similar to the 
value of the DA90 calibration, although the zero point can 
differ by  0.1--0.2 mag (see, e.g., the discussion in Salaris,
Cassisi \& Weiss~2002).

Therefore, there is a good general agreement on the way $M_{\rm bol}^{\rm TRGB}$
behaves as a function of age or metallicity. But distance determinations
using the RGB tip are mostly based on $I$- and $K$-band observations.
In the following we will separately discuss the calibrations in these two
bands.

\subsection{TRGB brightness in the $K$-band}

\begin{figure}
\includegraphics[width=8.3cm]{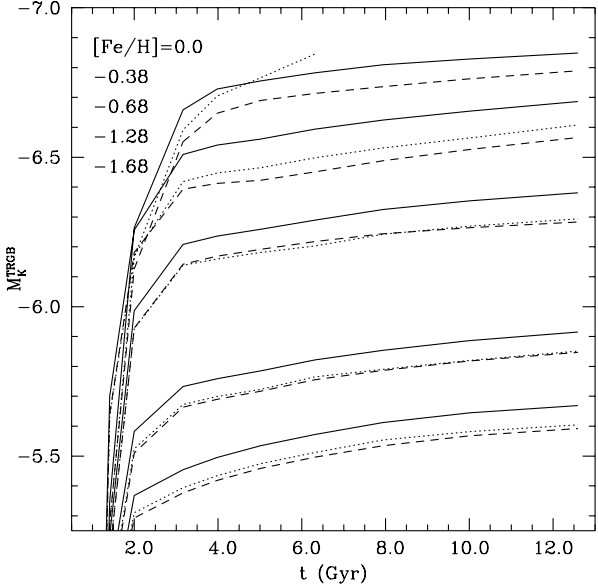}
\caption{$M_{K}$ of the TRGB as a function of age and [Fe/H]. 
The [Fe/H] values are listed in order of decreasing TRGB 
$K$-band brightness. The theoretical models are
by Girardi et al.~(2000), transformed to the $K$-band magnitude using
the theoretical bolometric corrections ($BC_K$) by Girardi et
al.~(2000 -- dashed line), the empirical $BC_K$ by Montegriffo et
al.~(1998 -- dotted line), and the combination of theoretical plus 
empirical $BC_K$
(empirical data for effective temperatures $T_{\rm eff} <$3500 K) 
presented in Girardi et al.~(2002 -- solid line)}
\label{kTRGB}
\end{figure}

Figure~\ref{kTRGB} displays the run of $M_{K}^{\rm TRGB}$ as a function
of age for various [Fe/H] values. Here we have used 
as a reference the theoretical models
by Girardi et al.~(2000), transformed to the $K$-band magnitude using
the theoretical bolometric corrections ($BC_K$) used by Girardi et
al.~(2000 -- dashed line), the empirical $BC_K$ by Montegriffo et
al.~(1998 -- dotted line), and the combination of theoretical plus 
empirical $BC_K$
(empirical data for effective temperatures $T_{\rm eff} <$3500 K) 
presented in Girardi et al.~(2002 -- solid line). 

It is easy to notice that in the
$K$-band the effect of metallicity is enhanced with respect to the
case of bolometric magnitudes. The general behaviour is the same
irrespective of the use of empirical or 
theoretical bolometric corrections that produces mainly zero point offsets.

In case of Galactic globular clusters
one finds semiempirically (e.g. Ferraro et al.~2000) that
$\Delta M_{K}^{\rm TRGB}/\Delta [Fe/H]\propto \sim -0.60$, i.e., the metallicity
dependence is a factor of $\sim$3 higher than for the bolometric
magnitudes, in agreement with the model predictions. 
This difference stems from the behaviour of $BC_K$, that is practically
independent of [Fe/H], and strongly affected by 
$T_{\rm eff}$ (see, e.g., Frogel, Persson \& Cohen~1981, 
Montegriffo et al.~1998) in the sense that $BC_K$ decreases strongly for
increasing $T_{\rm eff}$.
This produces fainter magnitudes for the hotter TRGBs, which are also
the more metal poor and fainter ones in $M_{\rm bol}$, hence the increased
dependence of $M_{K}^{\rm TRGB}$ on [Fe/H] when compared to $M_{\rm bol}^{\rm TRGB}$.  

One can also notice that, due to the behaviour of $BC_K$, there is a
non negligible dependence of $M_K^{\rm TRGB}$ on the age, in the sense
that a lower age at a given [Fe/H] mimics a lower metallicity at a
given age. 

In spite of the appeal due to the negligible dependence on reddening, 
$M_{K}^{\rm TRGB}$ is in principle prone to serious uncertainties when 
metallicity and age of the parent stellar population are uncertain.

\subsection{TRGB brightness in the $I$-band}

As shown by e.g., Bellazzini et al.~(2004) for Galactic
globular clusters, the TRGB
magnitude dependence on [Fe/H] decreases when moving from the $K$-band
to the $J$-band, i.e. $\diff M_{H}^{\rm TRGB}/\diff[Fe/H] \propto -$0.54  and 
$\diff M_{J}^{\rm TRGB}/\diff[Fe/H] \propto -$0.26.
In the $I$-band the trend of the bolometric correction $BC_I$ with [Fe/H]
tends to cancel the dependence of $M_{\rm bol}^{\rm TRGB}$ on
[Fe/H]. As an order of magnitude estimate we use 
$\diff M_{\rm bol}^{\rm TRGB}/\diff[Fe/H] \propto -$0.19 and recall that 
$M_{I}^{\rm TRGB}=M_{\rm bol}^{\rm TRGB}-BC_I$. DA90 found empirically that 
$\diff BC_I/\diff (V-I) \propto -0.243$ for RGB stars in a sample of 
Galactic globular clusters of various metallicity; 
Bellazzini, Ferraro \& Pancino~(2001)
showed empirically that the RGB colour in a sample of 
globulars scales with [Fe/H] as 
$\diff(V-I)/\diff$[Fe/H] $\propto$ 1.162  [Fe/H] + 2.472. 
This means that about [Fe/H]$= -$2.0 one has 
$\diff BC_I/\diff$[Fe/H] $\propto -$0.04
and about [Fe/H]$= -$0.7 one gets $\diff BC_I/\diff$[Fe/H] $\propto
-$0.40. Recalling the relationship between $M_{\rm bol}^{\rm TRGB}$ and
$M_I^{\rm TRGB}$ we get $\diff M_{I}^{\rm TRGB}/\diff[Fe/H] \propto -$0.15 about
[Fe/H]$= -$2.0 and $\diff M_{I}^{\rm TRGB}/\diff[Fe/H] \propto +$0.20 about 
[Fe/H]$= -$0.7, with an almost negligible dependence on [Fe/H] at
intermediate metallicities.
 
\begin{figure}
\includegraphics[width=8.3cm]{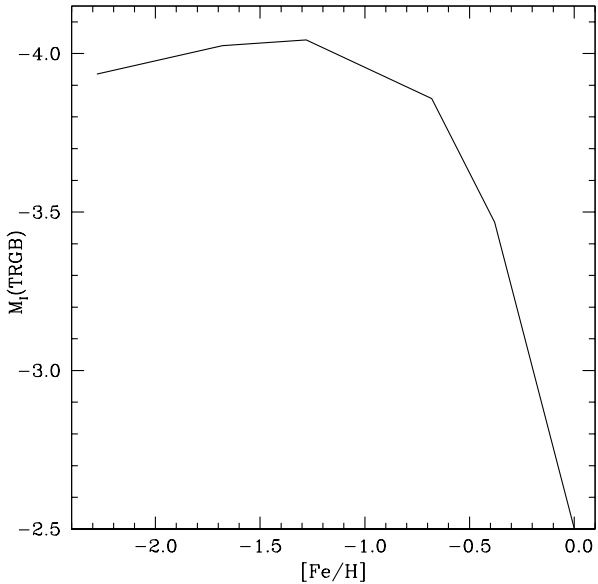}
\caption{$M_{I}$ of the TRGB as a function of [Fe/H], for
a set of 12.5 Gyr isochrones.}
\label{iTRGB}
\end{figure} 

This kind of relationship between $M_{I}^{\rm TRGB}$ and [Fe/H] is 
matched by theoretical models (see e.g., Salaris \&
Cassisi~1998) when considering ages typical of Galactic globular
clusters, on which this calibration is based. To illustrate further
this point, Fig.~\ref{iTRGB} displays the 
run of $M_{I}^{\rm TRGB}$ as a function
of [Fe/H] for a typical globular cluster age of 12.5 Gyr, 
using the theoretical models
by Girardi et al.~(2000) and the bolometric corrections 
by Girardi et al.~(2002). 
A comparison with Fig.~\ref{iTRGB} immediately
highlights the advantage of using the $I$-band instead of $K$ 
for old stellar populations: in fact, for [Fe/H] between -0.7 and
-2.0 $M_{I}^{\rm TRGB}$ varies less than 0.2 mag, as compared to the
$\sim1$ mag variation of $M_{K}^{\rm TRGB}$. 

The preferred TRGB distance methods are based therefore on $I$-band
observations. A very widely used technique is the one presented in
LFM93, based on Eq.~\ref{eq1} supplemented by the DA90 empirical
relationship 
\begin{equation}
BC_I=-0.243 \ (V-I) + 0.881
\label{eq2}
\end{equation}
and the following relationship between the RGB colour at $M_I=-3.5$ 
(denoted as $(V-I)_{0, -3.5}$) and [Fe/H]
\begin{equation}
{\rm [Fe/H]}=-12.64 + 12.6 \ (V\!-\!I)_{0, -3.5} - 3.3 \ (V\!-\!I)_{0, -3.5}^2.
\label{eq3}
\end{equation}

The LFM93 method works as follows. The apparent $I$-band magnitude of 
the TRGB is determined from the position of the 
luminosity function discontinuity, 
and after an estimate of the stellar
population reddening, one determines the intrinsic
colours of the observed RGB stars.
A preliminary distance modulus is then fixed and employed to determine a
preliminary metallicity by measuring $(V-I)_{0, -3.5}$ and using
Eq.~\ref{eq3}.
With this preliminar [Fe/H] estimate, one applies Eq.~\ref{eq1} to determine 
$M_{\rm bol}^{\rm TRGB}$, and Eq.~\ref{eq2} to determine $BC_I$ from the intrinsic 
mean colour of TRGB stars. With these two quantities one 
obtains a second approximation to the real distance modulus from 
\begin{equation}
(m-M)_0=I_{0, TRGB}+BC_I-M_{\rm bol}^{\rm TRGB}.
\label{eq4}
\end{equation}
This procedure is
then iterated until convergency is obtained. 
Since the almost vertical nature of the RGB, and the weak overall
dependency of $M_{\rm bol}^{\rm TRGB}$-$BC_I$ (as discussed in case of 
Galactic globulars on which this calibration relies) on [Fe/H], 
convergency is achieved usually within 2-3 iterations.

Alternative approaches to TRGB distance determinations 
either adopt theoretical $BC_I$ and $M_{\rm bol}^{\rm TRGB}$ values from
model atmosphere and stellar evolution computations to determine 
directly a theoretical
relationship $M_{I}^{\rm TRGB}$-[Fe/H] (as in Salaris \& Cassisi~1998) for
globular cluster-like ages, or
derive empirical $M_{I}^{\rm TRGB}$-[Fe/H] relationships by detecting the
TRGB in Galactic globular clusters with (empirically) known
distances and metallicities (e.g. Bellazzini, Ferraro \& Pancino~2001, 
Bellazzini et al.~2004).

\section{Application to composite stellar populations} 

As discussed extensively in the previous section, all TRGB
absolute magnitude calibrations applied for determining 
extragalactic distances are based on Galactic globular cluster
stars (observations, models or both). 
Any difference between the calibrating stellar populations and 
the observed ones may have an impact on the inferred TRGB
distances. As we will demonstrate shortly, {\em the presence of a well
populated RGB in a generic CMD does not automatically imply that the
stellar population under scrutiny is equivalent to Galactic globular
cluster populations}.
In case of spiral galaxies, observing far in the halo will very likely
minimize population differences, but when working with,
e.g., irregular galaxies (like many of the components of the 
Local Group) there is not 
a clear-cut distinction between the homogeneously old stellar 
component and younger ones.

In the following we will discuss these population effects on the 
TRGB distances to the LMC
and SMC by using population synthesis simulations of the galaxy
stellar populations, in the same vein as in Girardi \& Salaris~(2001).

\subsection{LMC}

\subsubsection{The simulated CMD}

\begin{figure}
\includegraphics[width=8.3cm]{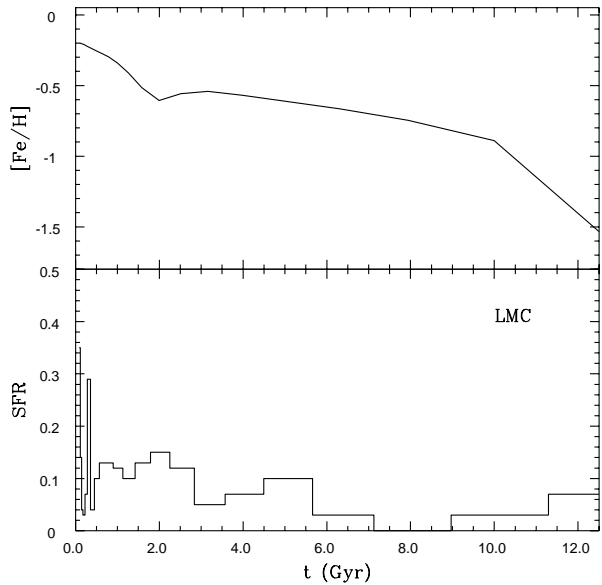}
\caption{SFR and AMR adopted for our LMC simulation. 
The SFR (bottom panel) is the one derived by Holtzman et al. 
(1999, his figure 4) from HST photometry of a field in the LMC bar. 
The AMR (top panel) comes from the bursting chemical evolution model of 
Pagel \& Tautvaisiene (1998), and agrees well with AMRs derived from 
observations of LMC star clusters (see e.g. Hill et al.~2000).}
\label{LMC1}
\end{figure}

We have computed synthetic CMDs of the LMC stellar populations using
the Star Formation History (SFH) and Age Metallicity Relation (AMR)
displayed in Fig.~\ref{LMC1}, which are typical of the LMC bar stellar
populations (see discussion in Girardi \& Salaris~2001, Salaris,
Percival \& Girardi~2003).
The synthetic CMDs have been computed 
using the same code employed by Girardi \& Salaris~(2001), 
and we made use of the stellar models by
Girardi et al.~(2000) and colour transformations by Girardi et al.~(2002). 
We have included in the simulation a Gaussian 1$\sigma$ 
photometric error of 0.02 mag 
(typical error at the magnitudes of LMC and SMC TRGB, see, e.g., the 
discussion in Barker et al.~2004), and a Gaussian 1$\sigma$ spread of 
0.1~dex around the [Fe/H] values displayed in fig.~\ref{LMC1}, 
but the following results are 
completely independent of the photometric errors and [Fe/H] dispersion.

The top panel of Fig.~\ref{LMC2} shows the $M_I-(V-I)$ CMD of 
the synthetic LMC RGB population. It is evident, in this figure,
the position of the TRGB at about $M_I\sim -4$.

\subsubsection{The apparent discrepancy in mean [Fe/H]}

\begin{figure}
\includegraphics[width=8.3cm]{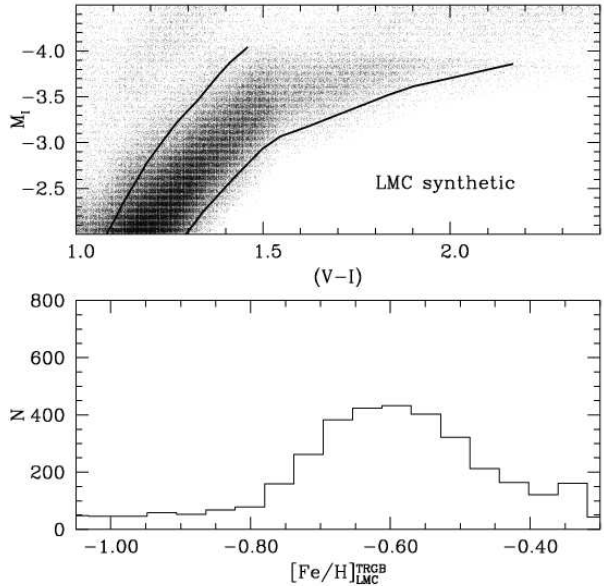}
\caption{Simulation of the LMC bar population using the SFR and
AMR of Fig.~\protect\ref{LMC1}. The top panel shows the 
simulated $M_I$ vs. $V-I$ CMD detailing the upper part of the RGB
(dots), and compares it with two globular cluster isochrones for  
[Fe/H] equal to $-$1.5 and $-$0.9, respectively (continuous thick
lines). The bottom panel shows 
the metallicity distribution of the simulated upper RGB stars.}
\label{LMC2}
\end{figure}

Overplotted in Fig.~\ref{LMC2} are the RGBs of two 
globular cluster isochrones (age of 12.5 Gyr) with the labelled 
[Fe/H] of $-$0.9 and $-$1.5.  They delimit the bulk of
RGB stars found in the simulation. By comparing the
$(V-I)$ colours of the globular cluster isochrones and 
the LMC population, we can derive, in a way similar to the
use of Eq.~\ref{eq3}, the mean [Fe/H] value for the RGB stars
in the LMC.
 
A first fundamental point has to be noticed at this stage. 
Our synthetic LMC population has been computed using the scaled solar
models by Girardi et al.~(2000); the LMC 
chemical evolution models by 
Pagel \& Tautvaisiene~(1998) and the spectroscopic observations by
Hill et al.~(2000) and Smith et al.~(2002) show that at low
metallicity the $\alpha$-elements (i.e. O, Ne, Mg, Si, S, Ca, Ti) 
are mildly enhanced with respect to Fe, i.e.,
[$\alpha$/Fe]$\sim$0.1--0.2. However, the relevant [Fe/H] range 
of the RGB stars will turn out to be -- as thoroughly discussed in the following
-- such that [$\alpha$/Fe]$\sim$0 according
to the results mentioned before, and our use of scaled solar models
for the simulation of LMC red giants is fully justified.

The Galactic globular cluster RGB isochrones employed to
determine the photometric metallicity of our LMC synthetic 
population, following the LFM93 method, must however take into account
the fact that [$\alpha$/Fe]$\sim$0.3 in globular cluster stars 
(see, e.g., Salaris, Chieffi \& Straniero~1993, Carney~1996 
and references therein).
For metallicities typical of the Galactic halo $\alpha$-enhanced 
theoretical stellar models are well reproduced by scaled solar ones 
with the same total metallicity [M/H] (Salaris et al.~1993), 
and therefore the globular cluster isochrones displayed in 
Fig.~\ref{LMC2} are the scaled solar ones by Girardi et al.~(2000) 
with the appropriate choice of the total metallicity that mimics 
[$\alpha$/Fe]=0.3, i.e. [M/H]=[Fe/H]+0.2 (for a scaled solar 
mixture [M/H]=[Fe/H]).

The mean [Fe/H] obtained comparing the mean colour 
of our synthetic LMC population at $M_I=-$3.5 
(the reference magnitude used in the LFM93 method)
with the colours of individual globular cluster 
isochrones, of various [Fe/H] and at the same $M_I$, 
is [Fe/H]=$-$1.08. This is in very good 
agreement with values obtained by LFM93 and Sakai et al.~(2000) 
from $VI$ LMC observations. We also redetermined 
the mean [Fe/H] of LMC giants employing photometry of the 
LMC OGLE field 6 (Udalski et al. 2000) and the method by LFM93. 
After correcting for the effect of extinction using the reddening maps
by Udalski et al.~(1999) we obtained a mean [Fe/H]=$-$1.15.

This [Fe/H] value obtained from the synthetic CMD and in agreement 
with estimates from observed CMDs is however greatly discrepant from 
the results of spectroscopic observations of 
bright RGB stars in the LMC bar published by Cole et al.~(2000), 
who found a mean [Fe/H]=$-$0.60.
It is important to notice that if we apply the LFM93 technique and 
calibrations, this [Fe/H] difference, taken at face value, 
would produce a systematic change in the distance modulus 
by $\sim$0.10 mag, because $BC_I$ in Eq.~\ref{eq4} is fixed by 
the observed $(V-I)$ colour, and $M_{\rm bol}^{\rm TRGB}$ changes by 
0.10 mag for a 0.5 dex change of [Fe/H]. 

The bottom panel of Fig.~\ref{LMC2} shows the [Fe/H] distribution 
for the stars in our simulation that 
populates the upper RGB. Here we find a mean [Fe/H]=$-$0.61, 
in excellent agreement with the spectroscopic result of $-0.60$. 
This means that also the synthetic LMC population displays the 
$\sim$0.5 dex discrepancy between the RGB 
[Fe/H] inferred from the RGB colour and the true value 
(the spectroscopic one in case of real LMC stars).

\begin{figure}
\includegraphics[width=8.3cm]{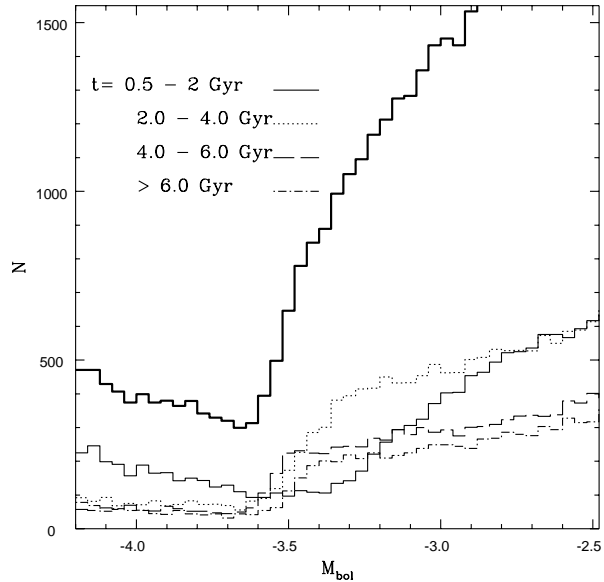}
\caption{Luminosity functions for all red stars ($V-I>1$) in the
LMC simulation (heavy line), and separated in four age bins 
(light lines). Notice that the TRGB discontinuity, detectable at 
about $M_{bol}=-3.6$, is mainly caused by stars in the $2-4$ and $4-6$ age 
bins.}
\label{LMC3}
\end{figure}

The reason for this discrepancy is twofold. 
The first cause is related to the age distribution of LMC stars 
compared to the age of globular clusters.
Figure~\ref{LMC3} displays the binned luminosity 
function (bins 0.04 mag wide) of all bright red stars 
($V-I>$1.0) in the LMC\footnote{We display the luminosity function in the
bolometric magnitude since the LFM93 method employs a theoretical
calibration of $M_{\rm bol}^{TRGB}$ as the standard candle of choice. Identical
results are obtained from the luminosity function in either 
the $I$- or $K$-band.} and the 
partial luminosity functions for different age ranges.
The discontinuity corresponding to the TRGB is clearly 
visible, and for what follows the 
technique used to identify the location of the discontinuity 
is completely irrelevant.

The mean age of bright RGB stars is $\sim$4 Gyr, 
much lower than the age (12--13 Gyr, see e.g. 
Salaris \& Weiss~2002) of the Galactic globular clusters 
on which the RGB colour-[Fe/H] relationship is calibrated.
Although the RGB colour is weakly sensitive to the age of 
the parent stellar population, 
an age difference of $\sim$9 Gyr has an appreciable influence, 
in the sense that 
the younger LMC population is bluer than the 
globular cluster one at fixed chemical composition.
If we shift the mean colour of the LMC RGB to the red, to account
for the effect of a 9~Gyr age difference with respect to a typical globular
clusters RGB, we would determine a photometric [Fe/H] value that is $\sim$0.3 dex
lower than the intrinsic one.

After the correction for the age effect there are still 
0.2 dex of mismatch in [Fe/H] to be accounted for. This remaining
difference is related to the different metal mixture 
of the calibrating globulars. Since a given [Fe/H] value coupled 
to [$\alpha$/Fe]$>$0 corresponds to 
a higher [M/H] than for a scaled solar mixture, 
RGB stars in the LMC appear to be bluer 
than the counterpart in globulars with the same [Fe/H]. 
This causes an underestimate of the LMC 
stars [Fe/H] of about 0.2 dex if [$\alpha$/Fe]$\sim$0.3 for 
Galactic globular clusters, and in this way the full difference
between photometric and real [Fe/H] is explained.
 
Let us for a moment suppose that the underestimate of [Fe/H] 
for the LMC RGB is due only to this latter effect, 
and denote with [Fe/H]$'$ the value obtained from Eq.~\ref{eq3} 
increased by 0.2 dex. We know that $M_{\rm bol}^{\rm TRGB}$ is determined 
by the total stellar metallicity [M/H] (e.g. Salaris et al.~1993), 
and that the calibration given by Eq.~\ref{eq1} is based on 
$\alpha$-enhanced globular cluster stars, for which [M/H]=[Fe/H]+0.2. 
This means that the use of the photometric [Fe/H] given by 
Eq.~\ref{eq3} instead of [Fe/H]$'$ does provide the 
correct $M_{\rm bol}^{\rm TRGB}$ for the LMC stars, because at a given 
[Fe/H] their [M/H] is lower by 0.2 dex than the globular cluster 
counterpart, and this mismatch is compensated for by the use of the 
'wrong' photometric [Fe/H]. 
The bottom line is that, from the point of view of the expected absolute
magnitude of the TRGB for LMC stars, the 0.2 dex [Fe/H] mismatch due to
the $\alpha$-enhancement in Galactic globular clusters does not play a role in
the LFM93 method.

\begin{figure*}
\begin{minipage}{0.65\textwidth}
\includegraphics[width=\textwidth]{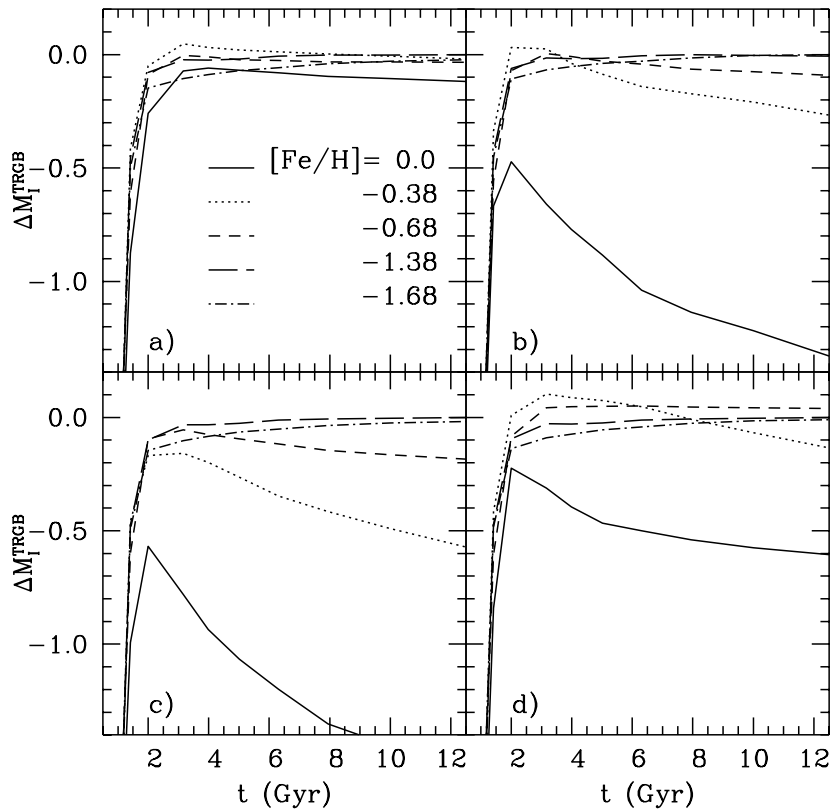}
\end{minipage}
\hfill
\begin{minipage}{0.33\textwidth}
\caption{Difference between the theoretical $M_I^{\rm TRGB}$ for t=12.5
Gyr, scaled solar [Fe/H]=$-$1.38 (corresponding to [Fe/H]$\sim -$1.6 
for the typical Galactic globular cluster
$\alpha$-enhancement) and the theoretical values predicted for the ages and
scaled solar [Fe/H] values displayed. The underlying
theoretical models are from Girardi et al.~(2000). Panel a), b), c)
and d) show, respectively, the results using the Yale transformations,
Westera et al.~(2002) transformations, Girardi et al.~(2002) 
transformations, and the transformations used in Girardi et al.~(2000).}
\label{diff}
\end{minipage}
\end{figure*} 

\subsubsection{Implications for the TRGB distance to the LMC
using the LFM93 method}

A few facts can be derived from the above discussion:

(1) The TRGB discontinuity (in $M_{\rm bol}$, but the same is true for 
the $I$- and $K$-band) in the
LMC is determined by stars with ages of the order of $\sim$4 Gyr. 
Older and younger RGB stars, although certainly present, are either 
too few (in the case of old stars), or do not contribute significantly 
to the detected TRGB discontinuity (in the case of stars younger than 
$2$~Gyr), or both. 
  
(2) Since most of the RGB stars in the LMC are intermediate-age, it is wrong 
to determine their [Fe/H] by simply comparing their $V-I$ colours 
with those of globular cluster giants.

(3) When using the LFM93 method in the LMC, a number of
additional steps have to be followed: Whereas $BC_I$ is fixed by 
the observations, $M_{\rm bol}^{\rm TRGB}$ {\em has to be corrected} 
for a 0.3 dex metallicity increase effect (that increases the
$M_{\rm bol}^{\rm TRGB}$ at fixed age) 
and for the age decrease with respect to the globular cluster ages (this
effect is however very small), 
as predicted by the theoretical models. Applying these corrections,
the LMC distance modulus increases by 0.05 mag with respect 
to results with the original LFM93 method and calibrations.

A particular remark should be made regarding item (2). As seen
in Fig.~4, in the LMC the SFR has been variable and presents a gap 
from 7 to 9 Gyr, that nearly coincides with the usual separation betwen
populations of intermediate ($\sim1$ to 7 Gyr) and old ages ($>$9 Gyr).
At intermediate ages, the mean SFR is just 3.1 times 
higher than the mean one at old ages. It may seem 
surprising that this moderate excess of the SFR at intermediate ages 
has caused the RGB to be {\em dominated} by $\sim$4 Gyr old stars,
with just a minor contribution from old ages (see Fig. 6). Actually,
there is no surprise in this result: The number of evolved stars of a given
kind and age is proportional to both the SFR at that age, and to their
intrinsic lifetime and birth rate. The birth rate is 
a function that strongly decreases with the stellar age, and this
would cause the predominance of intermediate-age stars even for 
the case of constant SFR.
This issue has been thoroughly discussed, in the case of red 
clump stars, by Girardi \& Salaris (2001); the situation for RGB stars
close to the TRGB is a very similar one. 

Therefore, the dominance of the intermediate-age RGB is not a
result specific to the LMC: it is the case for any galactic 
component in which the star formation, at intermediate ages, has been 
significant compared to old ages. This includes the 
discs of spirals, most of the irregulars,
and a significant fraction of the dwarf spheroidal galaxies.

\subsubsection{Implications for the TRGB distance to the LMC
using other methods}

Other methods usually consider a constant $M_I^{\rm TRGB}$ for 'metal poor'
stellar populations taken from calibrations based on Galactic
globulars. For example, Frayn \& Gilmore~(2003), Karechentsev et
al.~(2003), McConnachie et al.~(2004), use a constant $M_I^{\rm TRGB}$
value based on globular cluster calibrations, and apply it to the
observed RGB population of the parent galaxy.
Of course, in these cases we have again the assumption that the
RGB population, or at least the RGB stars that cause the detected
TRGB discontinuity, are old and metal-poor. As we have just shown,
this is not the case for the LMC bar population.

To check the bias in the LMC distance obtained with this
method one can use theoretical isochrones differentially, 
once we have identified --  by means of simulations like the one
of Fig.~\ref{LMC2} -- the typical ages and metallicities of TRGB stars.

First, we can directly compare $M_{I}^{\rm TRGB}$ obtained from the isochrones 
for a typical halo globular cluster metallicity [Fe/H]=$-$1.6 and an
age of 12.5 Gyr, with the $M_{I}^{\rm TRGB}$ obtained using the true
[Fe/H] and true age of the TRGB stars in the LMC. The difference between the two 
$M_{I}^{\rm TRGB}$ values, hereafter defined as
\begin{equation}
\Delta M_{I}^{\rm TRGB} = M_{I}^{\rm TRGB}(12.5Gyr,-1.6) - M_{I}^{\rm TRGB}(true) 
\end{equation}
provides the correction we have to apply to our
distance determination. We have performed this exercise,
based on the synthetic data of Fig.~\ref{LMC2}. 

Somewhat surprisingly, we find that the results depend in a non-negligible
way on the adopted set of transformations between the theoretical
$(M_{\rm bol}, T_{\rm eff})$ and the observational $(M_I, V-I)$ plane.
The results can be summarized as follows:
\begin{itemize}
\item Using Girardi et al.~(2000) transformations the LMC distance 
has to be increased by $0.07$ mag;
\item using Girardi et al.~(2002) transformations the LMC distance 
has to be decreased by $0.14$ mag;
\item using Yale transformations (Green~1988) the LMC distance is 
unchanged;
\item using Westera et al.~(2002) transformations the LMC distance is
unchanged.
\end{itemize}

Figure~\ref{diff} illustrates the behaviour of the quantity 
$\Delta M_I^{\rm TRGB}$ as a function of age and metallicity, for the
above-mentioned sets of transformations.  We recall that they 
are derived from libraries of stellar spectra comprising different 
metallicities. Since studying the 
reasons for the differences among these curves is beyond the scope 
of the present paper, we limit ourselves to a few comments.
In case of old metal poor populations, differences in $\Delta M_I^{\rm TRGB}$
are limited to $\approx$ 0.1 mag, but anyway cannot be
considered negligible. At high metallicities and old ages, as soon
as the $T_{\rm eff}$ value of TRGB stars becomes lower than $\sim 4000$~K, 
the discrepancies among the curves become dramatic. In fact,
the transformations in the range $T_{\rm eff}\la4000$~K are (1) subject to
larger variations due to the appearance of molecular bands
in stellar spectra, and (2) dealt with 
differently by different authors. A value $T_{\rm eff}=3500$~K is the 
minimum temperature for the ATLAS9 theoretical model atmospheres
(see, e.g., Kurucz~1993), which constitute the backbone of the transformations 
by Westera et al.~(2002) and Girardi et al.~(2000, 2002). Below this
$T_{\rm eff}$ (and to a certain extent also in the range $\sim3500-4000$ K),
the different transformations rely on sets of empirical spectra
and colour--$T_{\rm eff}$ relations, which are more prone to be implemented
in different ways. Therefore, it is no surprise that $\Delta M_I^{\rm TRGB}$
heavily depends on the adopted transformations in case of 
high metallicities.

It is also important to notice that the theoretical uncertainties in
$\Delta M_I^{\rm TRGB}$ cannot be avoided by using the
empirical relationship to
compute $BC_I$ employed by the LFM93 technique (see Eq.~\ref{eq2}), since
it has a non negligible 
1$\sigma$ scatter of 0.057 mag that DA90 attribute to observational errors. 
Moreover this empirical $BC_I$ calibration by DA90 does not reach 
the true mean [Fe/H] of the LMC RGB stars, its upper limit of validity
being [Fe/H]=$-$0.7 (the metallicity of 47~Tuc).

Could one avoid the uncertainties in the transformations, by
deriving a TRGB distance using infrared instead of $I$-band 
observations? As indicated by Fig.~\ref{kTRGB}, different
sets of transformations provide almost the same behaviour of
$M_K^{\rm TRGB}$ as a function of age and metallicity, thus suggesting 
that $\Delta M_K^{\rm TRGB}$ could be derived more reliably than 
$\Delta M_I^{\rm TRGB}$. The problem in this case is related to the
high dependence of $M_K^{\rm TRGB}$ on age and metallicity: 
The computation of the correction $\Delta M_K^{\rm TRGB}$ requires the
knowledge of age and metallicity distribution of RGB stars,
any error in these quantities are likely to produce substantial 
errors in $\Delta M_K^{\rm TRGB}$. Since the calibration of 
$M_K^{\rm TRGB}$ is based on globular clusters -- located
at one extreme of the age--metallicity plane allowed to RGB
populations -- biases are expected to be high.
The maximum bias is obtained when using the $K$-band magnitude of 
the TRGB coupled
to a semiempirical or theoretical $M_K^{\rm TRGB}$ calibration based on
globular clusters, and a calibration (e.g. Ferraro et al.~2000)
$(V-K)$-[Fe/H] or $(J-K)$-[Fe/H] obtained from globulars.
By using differentially the theoretical isochrones in infrared
passbands, the [Fe/H] and age biases discussed before would produce a bias of 
$\sim$0.2 mag (true distance larger than the value obtained from the
globular cluster calibration) in the TRGB distance, when considering a globular cluster age of 12.5
Gyr. In addition, for the $K$-band -- as mentioned before -- the
precise age of the Galactic globulars plays also a role.  

\subsection{SMC}

After discussing in detail the case of the LMC, we now examine the
SMC. Many of the considerations made for the LMC apply also to the 
SMC, and will not be repeated for the sake of conciseness.

\begin{figure}
\includegraphics[width=8.3cm]{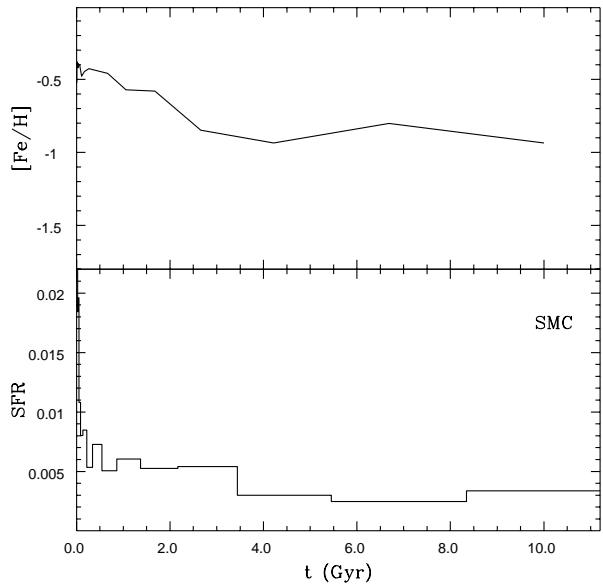}
\caption{SFR and AMR adopted for our SMC simulation. 
Both SFR and AMR are taken from Harris \& Zaritsky~(2004; 
their figures 7 and 12, respectively). The SFR represents the
global star formation in the SMC. The AMR agrees 
well with those derived from observations of SMC star clusters.}
\label{SMC1}
\end{figure} 

For our simulations, we select the 'global' SFR and AMR derived by
Harris \& Zaritski~(2004) from $UBVI$ photometry of the entire main
body of the SMC, and shown in Fig.~\ref{SMC1}. 
Their SFR is certainly the best available estimate 
for the SMC, whereas the AMR relation is in good agreement with 
those derived from observations of SMC star clusters.
Scaled solar models are again appropriate for the relevant
metallicities of the RGB stars, according to the discussion in 
Pagel \& Tautvaisiene~(1998 -- their Fig.~9 and discussion in Sect.~3).

\begin{figure}
\includegraphics[width=8.3cm]{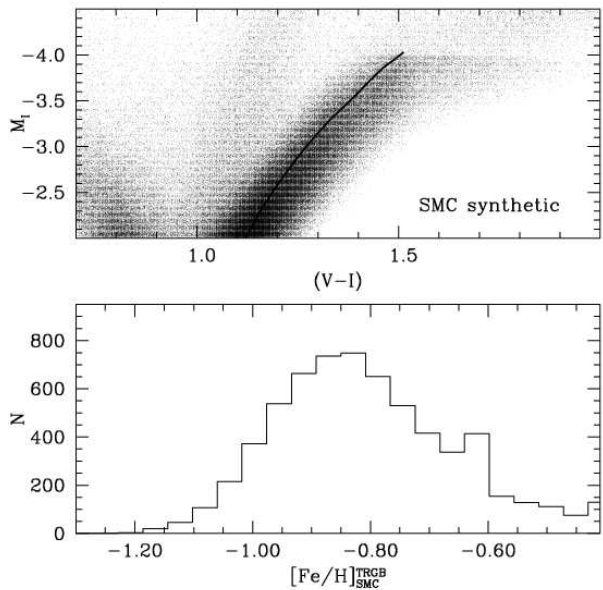}
\caption{The same as  Fig.~\ref{LMC2}, but for the
SMC, using the AFR and AMR of Fig.~\ref{SMC1}. The
overplotted globular cluster isochrone (thick continuous line) has [Fe/H]=$-$1.4.}
\label{SMC2}
\end{figure} 

The simulation is presented in Fig.~\ref{SMC2}. The 
photometric mean [Fe/H], as derived from comparing the RGB with those of
old isochrones and considering their $\alpha$-enhancement, is
[Fe/H]=$-$1.4. From OGLE-II photometry of SMC OGLE field 3, dereddened, 
we get a similar value, [Fe/H]=$-$1.5.

The true mean RGB metallicity, as derived from the
simulation itself, is [Fe/H]=$-$0.82. Again, we have a discrepancy of
0.6 dex between the two values of [Fe/H], that can be explained as follows:
0.2 dex are due to the $\alpha$-enhancement, 0.4 dex are 
due to the mean younger age of SMC RGB stars with respect to
old globular clusters. In fact, the mean age of bright RGB stars in 
the SMC is of just $\sim$3.7 Gyr. 

\begin{figure}
\includegraphics[width=8.3cm]{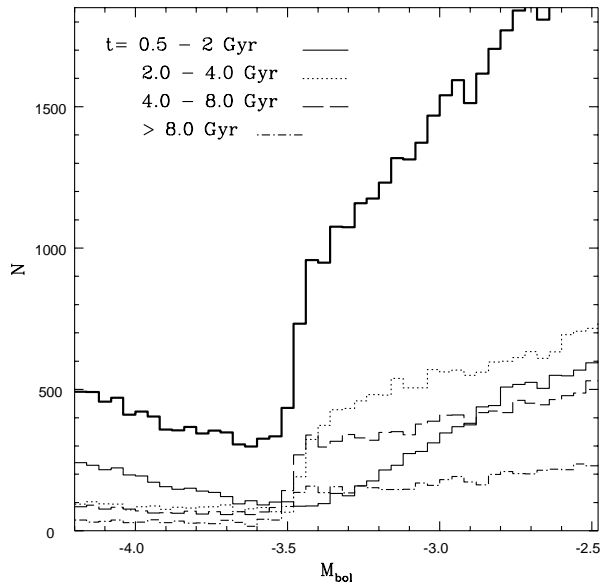}
\caption{The same as  Fig.~\protect\ref{LMC2}, but for the
SMC simulation. The age ranges for the older ages are slightly
different, to highlight the different age distribution of SMC stars
with respect to the LMC.}
\label{SMC3}
\end{figure}

The TRGB discontinuity is mainly determined by stars with 
ages $\sim$4 Gyr, and not by the oldest component with globular
cluster-like ages, as is evident by looking at the partial LFs depicted
in Fig.~\ref{SMC3}.

Using the LFM93 method, the effect of the [Fe/H] discrepancy 
is an increase of 0.07 mag for the SMC distance.

Using the 'constant $M_I^{\rm TRGB}$' methods, the results again depend 
on the selected set of transformations. Using Girardi et al~(2002) 
ones, the SMC distance has to be decreased by 0.04 mag.
With the other transformations tested in Fig.~\ref{diff}, 
the SMC distance is basically unchanged. In the $K$-band, the
$\Delta M_K^{\rm TRGB}$ amounts to $\sim$0.2 mag, for all the different
sets of transformations.

Comparing the LMC and SMC cases, we can conclude the following:
In both cases, the use of LFM93 will lead to a small underestimate
of their distance moduli, amounting to 0.05 for the LMC, and 0.07 mag 
for the SMC. These small errors are still lower than
the overall accuracy of the TRGB methods (defined by the uncertainty 
in the zero points, of $0.1-0.2$ mag), but are not completely 
negligible. Moreover, the errors seem to be systematic in the sense
of decreasing distances to the galaxies that have had 
significant star formation at intermediate-ages. Fortunately enough, 
there is no impact on the relative SMC--LMC distance (but one has
always to keep in mind that the $BC_I$ relationship by DA90 is used
outside its range of validity for the LMC TRGB).

When using the 'constant $M_I^{\rm TRGB}$' methods, instead, the results
depend on the set of transformations one uses to derive the 
quantity $\Delta M_I^{\rm TRGB}$. Using the Girardi et al. (2002) 
transformations, for instance, one derives that the straight use of
a constant $M_I^{\rm TRGB}$ value lead to errors in distance
moduli amounting to $-0.14$ mag for the LMC, and $-0.04$ mag for the
SMC. This would imply an error of 0.10 mag (true distance being
longer) in the derived 
relative SMC-LMC distance. This is a significant error, that may 
bias the calibration of other standard candles as a function of 
the metallicity. On the other hand, if one uses Westera et al. (2002) 
transformations to derive $\Delta M_I^{\rm TRGB}$, both SMC and LMC 
distance moduli remain unchanged, as well as their relative distance 
and the calibration of other standard candles on the Magellanic Clouds.

\subsection{LGS3}

\begin{figure}
\includegraphics[width=8.3cm]{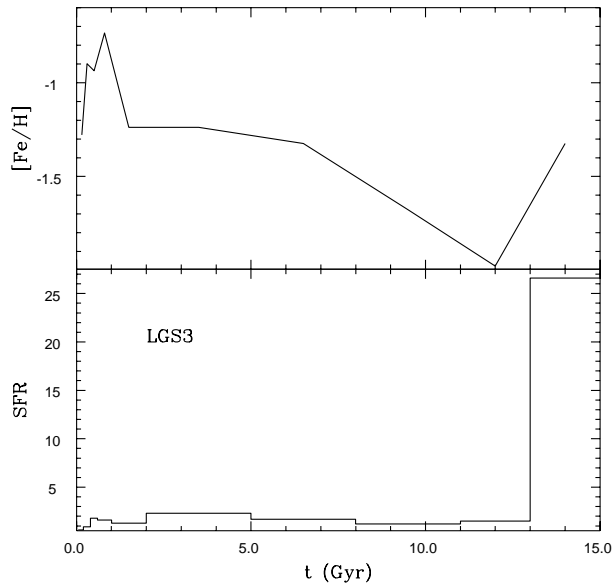}
\caption{SFR and AMR adopted for our LGS3 simulation, 
from Miller et al.~(2001).}
\label{LGS3a}
\end{figure}

As an example of galaxy where the globular cluster TRGB calibration is
appropriate, we have considered the case of LGS3. Miller et
al.~(2001) provide a global SFR and AMR for this galaxy, displayed in Fig.~\ref{LGS3a}. 
A main burst of star formation happened at the beginning of the
galaxy life, after which the SFR level has been very low and almost constant.
The spread around the mean AMR is also accounted for, following Miller
et al.~(2001) estimates.

\begin{figure}
\includegraphics[width=8.3cm]{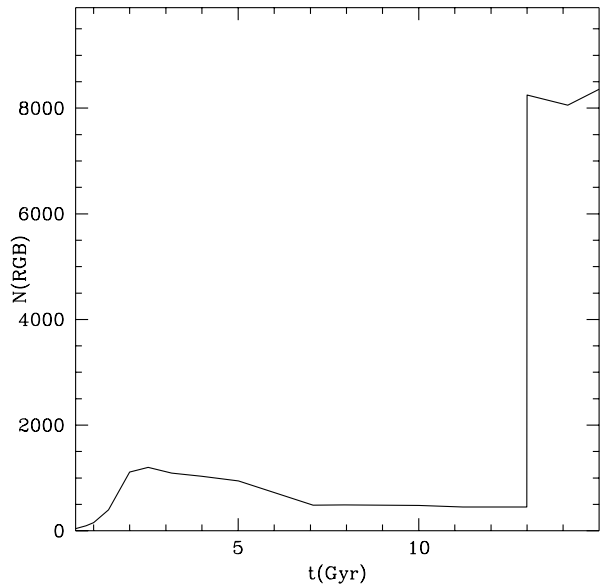}
\caption{Age distribution of the RGB stars obtained from the LGS3 simulation.}
\label{LGS3b}
\end{figure}

Figure~\ref{LGS3b} displays the age distribution of RGB stars obtained
from our simulation. The main component is made of the oldest objects
formed during the strong initial burst of star formation. RGB stars 
born after the initial burst amount only to about half of the oldest objects.  
Since the age and metallicity ([Fe/H]=$\sim -$1.3 for
the majority of LGS3 RGB objects) of the bulk of the LGS3 RGB stars 
is comparable to the globular
cluster counterparts, the standard calibrations of the TRGB absolute
magnitude can be applied with confidence to this galaxy.

More in detail, when employing the LFM93 method a 0.2 dex underestimate of [Fe/H]
is possible, if LGS3 stars formed all with a scaled solar metal
distribution (as assumed in our simulation). However, as seen before,
this does not affect the TRGB distances with LFM93 technique.
Also when employing the 'constant $M_I^{\rm TRGB}$' methods, the true  distance
is recovered.

The relative distance modulus LGS3-LMC is affected at the level of 0.05 mag by
the population effects discussed before when using the LFM93 method. 
The 'constant $M_I^{\rm TRGB}$' methods provide different results,
depending on the set of bolometric transformations adopted, the
maximum effect amounting to a significant 0.14 mag on the relative distance modulus.

\section{Conclusions} 
 
The previous analysis of the TRGB distances to the galaxies LMC, SMC and LGS3 
allowed us to point out a series of uncertainties affecting the current TRGB 
distance determinations and RGB photometric metallicity estimates 
for composite stellar populations.

All TRGB distance methods rest on the hypothesis that the 
RGB populations observed in external galaxies are as old as the stars
in Galactic globular clusters, on which the TRGB calibrations are based. 
This is however not always true, as
we have demonstrated for the specific cases of the LMC and SMC, whose RGB
populations have ages of the order of 3-4 Gyr, instead of 12-13 Gyr
typical of the Galactic globulars. This age difference (and eventual 
differences in the [$\alpha$/Fe] ratio between the Galactic globular
clusters and external galaxies) affects the TRGB distances and [Fe/H]
estimate of the parent RGB stars, as has been already pointed out by
Davidge~(2003), Rizzi et al.~(2003), Salaris et al.~(2003).
In case of LMC and SMC we found that [Fe/H] estimates based on 
the comparison of the RGB colour with the globular cluster counterparts are
too low by $\sim$0.5 dex due to the effect of both age ($\sim$0.3 dex) and
different heavy element distribution ($\sim$0.2 dex). These effects
explain the discrepancy between the photometric metallicity of the LMC
RGB determined from a comparison with the colours of globular cluster RGB stars,
and the spectroscopic measurements.

The younger age of a galaxy RGB
population with respect to globular cluster ages 
may induce a bias also in the $I$-band TRGB distance, whose extent is
strongly dependent on the adopted $BC_I$ scale. 
TRGB relative distances between Magellanic Clouds-type galaxies and
galaxies like LGS3 populated by mainly globular cluster-like stars 
can be appreciably affected by these population effects, up to 
$\sim$0.15 mag, depending on the $BC_I$ scale adopted. 
More in detail, the correction to apply to the LGS3--LMC
distance modulus ranges between $-0.05$ mag and $+0.14$ mag, whereas in case
of the LGS3--SMC distance modulus it ranges between $-$0.07 mag and
$+$0.04 mag. As for the distance modulus SMC--LMC, 
the correction due to population effect ranges from 0 mag up to +0.10 mag. 
 
The use of $K$-band TRGB calibrations -- the $K$-band bolometric correction
scale appears to be more secure as far as the trend with $T_{\rm eff}$ is
concerned -- does not help much, due the strong dependence of
$M_{K}^{\rm TRGB}$ on both [Fe/H] and age; biases of the order of 0.2 dex 
can be present in $K$-band TRGB distances 
when using calibrations of the TRGB method based on Galactic globular
cluster stars.

In order to isolate the most metal poor and oldest RGB component to which
apply safely the TRGB globular cluster calibration, one might try 
to apply a colour cut to the RGB data, i.e. selecting stars bluer than
a given $(V-I)$ value, as suggested by our referee. 
We have however verified in the specific cases of
the SMC and LMC that this procedure does not achieve in general the desired results. 
More in detail, we have divided the RGB with $-3.70 \leq M_I \leq -4.05$
into three colour intervals, i.e. $1.3 \leq (V-I) < 1.5$ (blue), $1.5 \leq
(V-I) < 1.7$ (middle) and $1.7 \leq (V-I) < 1.9$ (red) and studied the associated distributions
of age and [Fe/H] values.
For SMC stars we find that the oldest population is spread
between the red and middle interval, whereas for the LMC the oldest
population is located in the blue interval. This difference reflects
the AMR of the two galaxies. In case of the SMC the AMR
used in these simulations is approximately flat for ages down to
$\sim$ 3~Gyr; a
sizable fraction of RGB stars has therefore a very similar metallicity,
and the oldest objects are necessarily located towards the red side of the RGB.
For the LMC the oldest stars are also the most metal poor and
therefore they tend to cluster at the blue side of the RGB.
It is evident from this example that the same colour cut 
applied to different galaxies may imply very different distributions of
ages and [Fe/H] due to different star formation and chemical enrichment
histories; we have 
therefore to conclude that there is no general rule for
selecting a particular kind of RGB stellar population based on 
colour cuts along the RGB CMD.

As for the possibility to detect the TRGB of the oldest populations in
the LMC and SMC using the colour analysis described before, the
following has to be noticed. The oldest SMC population 
is generally superimposed to the
younger and more metal rich one, that determines the level of
the TRGB discontinuity. As for the LMC, the bluest RGB stars in a
narrow colour range ($1.3 \leq (V-I) < 1.5$) can
in principle provide a reasonably clean sample of an old globular
cluster-like population, but their CMD location overlaps with
a sizable number of bluer objects belonging to different evolutionary
phases of much younger ($\la0.5$~Gyr) populations, that can be 
clearly seen in the upper left part of Fig.~5. 
The large fraction of these non-RGB objects populating the
same colour bin may hamper a clear detection of the LMC old 
metal poor TRGB.

To conclude, our results show that the presence of a well developed RGB
in the Colour Magnitude Diagram of a stellar system with a complex SFR does
not guarantee that the RGB is populated by globular cluster-like red
giants, nor that its TRGB is determined by them; 
this means that {\em the TRGB method for distance determinations 
has to be applied with caution to all galaxies which present signatures
of intermediate-age stars.} 
A definitive assessment of the appropriate corrections for
population effects on TRGB distances has however to wait for a
substantial reduction of the uncertainties on the $BC_I$
scale for cold stars.

\section*{Acknowledgments} 
We warmly acknowledge the hospitality at the Max Planck Institut f\"ur
Astrophysik, where part of this work was carried out. We thank Santi
Cassisi for stimulating discussions and the referee, B. Madore, for useful
suggestions. The work by L.G. has been funded by COFIN 2002028935-003.

\label{lastpage} 
 

\begin{thebibliography}{99} 
\bibitem[]{} Barker M.~K., Sarajedini A., Harris J., 2004, ApJ, 606, 869 
\bibitem[]{} Bellazzini M., Ferraro F.~R., Pancino E., 2001, ApJ, 556, 635 
\bibitem[]{} Bellazzini M., Ferraro F.~R., Sollima A., Pancino E., 
	Origlia L., 2004, A\&A 424, 199
\bibitem[]{} Carney, B.W., 1996, PASP, 108, 900
\bibitem[]{} Cioni M.-R.~L., van der Marel R.~P., Loup C., Habing H.~J., 
	2000, A\&A, 359, 601 
\bibitem[]{} Da Costa G.~S., Armandroff T.~E., 1990, AJ, 100, 162 
\bibitem[]{} Davidge T.~J., 2003, PASP, 115, 635 
\bibitem[]{} Ferraro F.~R., Montegriffo P., Origlia L., 
	Fusi Pecci F., 2000, AJ, 119, 1282 
\bibitem[]{} Frayn C.~M., Gilmore G.~F., 2003, MNRAS, 339, 887 
\bibitem[]{} Frogel, J.~A., Persson, S.~E., Cohen, J.~G, 1981, ApJ,
246, 842
\bibitem[]{} Girardi L., Salaris M., 2001, MNRAS, 323, 109 
\bibitem[]{} Girardi L., Bertelli G., Bressan A., Chiosi C., 
	Groenewegen M.~A.~T., Marigo P., Salasnich B., Weiss A., 2002, 
	A\&A, 391, 195 
\bibitem[]{} Girardi L., Bressan A., Bertelli G., 
	Chiosi C., 2000, A\&AS, 141, 371 
\bibitem[]{} Green, E.M., 1988, in ``Calibration of Stellar ages'',
A.G. Davis Philip ed. (L. Davis Press, Schenectady) p.~81
\bibitem[]{} Harris J., Zaritsky D., 2004, AJ, 127, 1531
\bibitem[]{} Harris W.~E., Durrell P.~R., Pierce M.~J., Secker J., 1998, 
	Nature, 395, 45 
\bibitem[]{} Hill V., Fran{\c c}ois P., Spite M., Primas F., Spite F., 
	2000, A\&A, 364, L19 
\bibitem[]{} Holtzman J.A., Gallagher J.S., Cole A.A., Mould J.R.,  
	Grillmair C.J., et al., 1999, AJ 118, 2262
\bibitem[]{} Jerjen H., Rejkuba M., 2001, A\&A, 371, 487 
\bibitem[]{} Karachentsev I.~D., et al., 2003, A\&A, 404, 93 
\bibitem[]{} Kippenhahn, R., Weigert, A., 1990, 'Stellar structure and
evolution', Springer-Verlag (Berlin, Heidelberg)
\bibitem[]{} Kurucz, R.~L., 1993, in ``The Stellar Populations of
Galaxies'', B. barbuy, A. Renzini eds. (Kluwer, Dordrecht), p.~225
\bibitem[]{} Lee M.~G., Freedman W.~L., Madore B.~F., 1993, ApJ, 417, 553 
\bibitem[]{} M\'endez, B., Davis, M., Moustakas, J., Newman, J.,
	Madore, B.F. \& Freedman, W.L. 2002, ApJ, 124, 213
\bibitem[]{} McConnachie A.~W., Irwin M.~J., Ferguson 
	A.~M.~N., Ibata R.~A., Lewis G.~F., Tanvir N., 2004, MNRAS, 350, 243 
\bibitem[]{} Miller B.~W., Dolphin A.~E., Lee M.~G., Kim S.~C., Hodge P., 
	2001, ApJ, 562, 713 
\bibitem[]{} Montegriffo, P., Ferraro, F.~R., Origlia, L., 
        Fusi Pecci, F., 1998, MNRAS, 297, 872
\bibitem[]{} Pagel B.E.J., Tautvaisiene G., 1998, MNRAS 299, 535
\bibitem[]{} Pietrzy{\' n}ski G., Gieren W., Udalski A., 2003, AJ, 125, 2494 
\bibitem[]{} Rizzi L., Held E.~V., Bertelli G., Saviane I., 2003, 
	ApJ, 589, L85 
\bibitem[]{} Sakai S., Madore B.~F., Freedman W.~L., 1996, ApJ, 461, 713
\bibitem[]{} Sakai S., Madore B.~F., Freedman W.~L., Lauer T.~R., 
	Ajhar E.~A., Baum W.~A., 1997, ApJ, 478, 49 
\bibitem[]{} Sakai S., Ferrarese L., Kennicutt R.~C., Saha A., 2004, 
	ApJ, 608, 42 
\bibitem[]{} Salaris M., Cassisi S., 1997, MNRAS, 289, 406 
\bibitem[]{} Salaris M., Cassisi S., 1998, MNRAS, 298, 166 
\bibitem[]{} Salaris, M., Weiss, A., 2002, A\&A, 388, 492
\bibitem[]{} Salaris, M., Cassisi, S., Weiss, A., 2002, PASP, 114, 375
\bibitem[]{} Salaris, M., Chieffi, A., Straniero, O., 1993, ApJ, 414, 580
\bibitem[]{} Salaris, M., Percival, S., Girardi, L., 2003, MNRAS, 345, 1030
\bibitem[]{} Sarajedini A., et al., 2002, ApJ, 567, 915 
\bibitem[]{} Smith V.~V., et al., 2002, ApJ, 124, 3241 
\bibitem[]{} Sweigart, A.~V.,Gross, P.~G., 1978, ApJS, 36, 405
\bibitem[]{} Sweigart A.~V., Greggio L., Renzini A., 1990, ApJ, 364, 527 
\bibitem[]{} Udalski A., Soszynski I., Szymanski M., Kubiak M.,
Pietrzynski G., Wozniak P., 1999, AcA, 49, 223
\bibitem[]{} Udalski A., Szymanski M., Kubiak M., Pietrzynski G., 
	Soszynski I., Wozniak P., Zebrun K., 2000, AcA, 50, 307 
\bibitem[]{} Westera, P., Lejeune, T., Buser, R., Cuisinier, F., 
Bruzual, G., 2002, A\&A, 381, 524
\end{thebibliography}
\end{document}